\documentclass[lettersize,journal]{IEEEtran}
\usepackage{amsmath,amsfonts}
\pdfoutput=1
\usepackage{algorithm}
\usepackage{array}
\usepackage[caption=false,font=normalsize,labelfont=sf,textfont=sf]{subfig}
\usepackage{textcomp}
\usepackage{stfloats}
\usepackage{url}
\usepackage{verbatim}
\usepackage{graphicx}
\usepackage{cite}
\usepackage{float}
\usepackage{marvosym}
\usepackage{multicol}
\usepackage{multirow}
\usepackage[english]{babel}

\newtheorem{definition}{Definition}
\newtheorem{assumption}{Assumption}
\usepackage{algpseudocode}


\usepackage{bbding}

\usepackage{booktabs}
\usepackage{colortbl}
\usepackage{pifont}
\usepackage[svgnames]{xcolor}
\definecolor{mygrey}{RGB}{230,230,230}
\definecolor{myred}{RGB}{218 244 246}
\definecolor{lightorange}{RGB}{218 244 246} 
\definecolor{mediumorange}{RGB}{182 219 222} 
\definecolor{darkorange}{RGB}{145 192 196} 
\usepackage{xcolor}
\newcommand{\responseblack}[1]{\textcolor{black}{#1}}

\begin{document}

\title{Federated Consistency- and Complementarity-aware Consensus-enhanced Recommendation}

\author{Yunqi Mi, Boyang Yan, Guoshuai Zhao\textsuperscript{\Letter}, ~\IEEEmembership{Member,~IEEE,} Jialie Shen, ~\IEEEmembership{Senior Member,~IEEE,} and Xueming Qian.

\thanks{\textsuperscript{\Letter} Corresponding author.}

\thanks{Yunqi Mi, Boyang Yan, and Guoshuai Zhao are with the School of Software Engineering, Xi’an Jiaotong University, Xi’an
710049, China (e-mail: miyunqi@stu.xjtu.edu.cn; yanboyang@stu.xjtu.edu.cn;
guoshuai.zhao@xjtu.edu.cn\responseblack{).}}
\thanks{Jialie Shen is with City, University of London, U.K.(e-mail: jerry.shen@city.ac.uk).}
\thanks{Xueming Qian is with the Ministry of Education Key Laboratory for Intelligent Networks and Network Security, the School of Information and Communication Engineering, and SMILES LAB, Xi’an Jiaotong University, Xi’an
710049, China (e-mail: qianxm@mail.xjtu.edu.cn).}}

\markboth{IEEE Transactions on Knowledge and Data Engineering}%
{Shell \MakeLowercase{\textit{et al.}}: A Sample Article Using IEEEtran.cls for IEEE Journals}

\maketitle

\begin{abstract}
Personalized federated recommendation system (FedRec) has gained significant attention for its ability to preserve privacy in delivering tailored recommendations.
To alleviate the statistical heterogeneity challenges among clients and improve personalization, decoupling item embeddings into the server and client-specific views has become a promising way. Among them, the global item embedding table serves as a consensus representation that integrates and reflects the collective patterns across all clients.
However, the inherent sparsity and high uniformity of interaction data from massive-scale clients results in degraded consensus and insufficient decoupling, reducing consensus's utility.
To this end, we propose a \textbf{Fed}erated \textbf{C}onsistency- and \textbf{C}omplementarity-aware \textbf{C}onsensus-enhanced \textbf{R}ecommendation (Fed3CR) method for personalized FedRec.
To improve the efficiency of the utilization of consensus, we propose an \textbf{A}daptive \textbf{C}onsensus \textbf{E}nhancement (ACE) strategy to learn the relationship between global and client-specific item embeddings. It enables the client to adaptively enhance specific information in the consensus, transforming it into a form that best suits itself.
To improve the quality of decoupling, we propose a \textbf{C}onsistency- and \textbf{C}omplementarity-aware \textbf{O}ptimization (C2O) strategy, which helps to learn more effective and complementary representations.
Notably, our proposed Fed3CR is a plug-and-play method, which can be integrated with other FedRec methods to improve its performance.
Extensive experiments on four real-world datasets represent the superior performance of Fed3CR.

\end{abstract}

\begin{IEEEkeywords}
Recommender system, federated learning, multi-view learning.
\end{IEEEkeywords}

\section{Introduction}
\IEEEPARstart{P}{ersonalized} federated recommender systems (FedRec) have gained substantial traction in both industry and academia due to their dual capability of delivering tailored recommendations while maintaining user privacy~\cite{ali2024hidattack, ye2023adaptive, intro2_ua, intro1_privaterec}.
These systems leverage a distributed architecture where clients (\textit{i.e.}, users) collaboratively train a recommender model without directly exposing their private interactions or sensitive information. However, the effectiveness of FedRec is often challenged by the inherent sparsity and high uniformity of interaction data among clients. This results in significant statistical heterogeneity and, in some cases, distribution drift across the clients~\cite{intro5_co, intro6_meta, intro7_survey}.
To address these challenges, a promising approach involves decoupling the global and personal perspectives of item embeddings, managing them separately at the server and client sides~\cite{pfedrec, gpfedrec, fedrap}.
In this framework, clients use their private data to learn global and personal item embedding representations. The personal embeddings remain local on the client, while the global embeddings are shared with the server. The server aggregates the global embeddings from all clients to generate a unified global representation, which is subsequently used to accommodate diverse client distributions. Consequently, the global item embedding table serves as a \textbf{consensus} representation that integrates and reflects the collective patterns across all clients, enhancing the robustness and adaptability of the recommendation system.

\begin{figure}[t]
    \centering
    \subfloat[{Personalized FedRec with degraded consensus}\label{intro.sub.1}]{
        \includegraphics[width=8cm,height = 3cm]{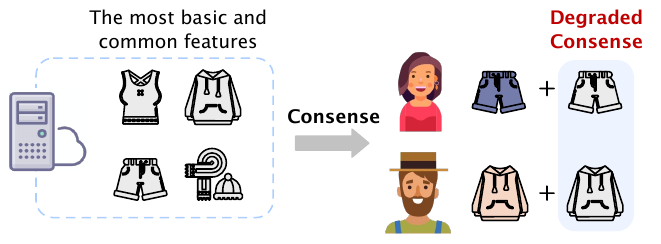}
        
        }
        \\
    \subfloat[{Fed3CR with enhanced consensus}\label{intro.sub.2}]{
        
        \includegraphics[width=8cm,height = 3cm]{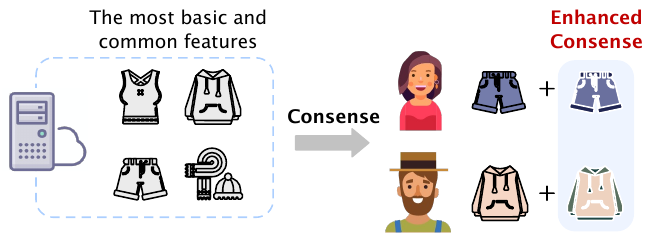}
    }

\vspace{-0.1cm}

\caption{The main difference between Fed3CR and other personalized FedRec approaches is the adaptive global item embedding enhancement for the client. It amplifies specific signals in the consensus for the client by identifying the relationship between personal embedding and the global consensus, and obtains the content that best suits the client.}
\vspace{-0.3cm}
\label{fig:intro}
\end{figure}

However, the utilization of consensus in clients is less efficient due to the following issues:
1) \textbf{Degraded consensus.} 
In FedRec, due to the high heterogeneity of a large number of clients (hundreds of thousands or even millions of clients \cite{clientnum1, clientnum2}), the server can only aggregate a degraded consensus to accommodate all clients.
Specifically, it is challenging for the server to produce knowledge that fits all distributions and contains sufficient information because the gradients cancel each other out when the client distributions vary widely.
Besides, the degraded consensus is so different from the distribution in the client that it is difficult to provide effective information for feature fusion, as shown in Figure~\ref{intro.sub.1}.
Moreover, the degree of degradation of the consensus varies for each client, and the more unique the personal distribution of clients, the harder it is for the aggregated consensus to provide valid information.
2) \textbf{Insufficient decoupling and fusing.}
Compared to other multi-view learning tasks~\cite{kdd_consistent1, consistent2, consistent3,fang2023comprehensive}, it is difficult to maintain consistent semantic information between global and personal perspectives due to the high heterogeneity among clients. This leads to an inevitable semantic mismatch issue, reducing the representational ability of the fused embeddings.
In addition, maintaining the complementarity of global and personal perspectives is necessary to improve the expressiveness of the fused item embeddings~\cite{fedrap, kdd_complementary1}.
FedRAP~\cite{fedrap} offers an insightful solution by pushing specific item embeddings away from shared item embeddings to achieve complementarity.
However, due to the utility of distance-based constraints (\textit{i.e.},~Euclidean distance), it still captures shared features, leading to information redundancy~\cite{orth1, orth2, orth3}.
Neglecting consistency and complementarity causes insufficient decoupling between global and personal perspectives, leading to negative enhancement and reducing the utility.

To handle these challenges, we propose a  \textbf{Fed}erated \textbf{C}onsistency- and   \textbf{C}omplementarity-aware \textbf{C}onsensus-enhanced (\textbf{Fed3CR}) approach for personalized FedRec.
In this paper, we first discuss the consensus degradation issue in FedRec in detail and provide some mathematical analysis.
We aim to improve the efficiency of consensus utilization in the client, and view the server as a teacher who imparts general consensus without personalized instruction.
The client uses the \textbf{A}daptive \textbf{C}onsensus \textbf{E}nhancement (ACE) module to analyze the relationship of global and personal item embeddings, learning how to enhance specific information in the consensus. It adaptively transforms consensus into a form that best suits the client, as shown in Figure~\ref{intro.sub.2}.
Furthermore, to achieve sufficient and efficient decoupling of global and local item embedding, we define consistency and complementarity in FedRec as follows:

\begin{definition}
    \textbf{Consistency : }The semantic spaces of items between the server and clients should remain consistent. This means that the relationships among items observed in the global view should align with those in the client-specific personal view. 
\end{definition}
\begin{definition}
    \textbf{Complementarity : }Global and personal item embeddings should encapsulate distinct but complementary aspects of the items. Together, they should form a holistic and enriched representation that leverages both shared global patterns and individualized client-specific nuances.
\end{definition}
Then, we propose a \textbf{C}onsistency- and \textbf{C}omplementarity-aware \textbf{O}ptimization (C2O) strategy. It comprehensively explores the consistency and complementarity between the global and personal views, effectively preventing feature redundancy and mismatches.
Specifically, we focus on the relative consistency between semantic spaces rather than absolute similarity and propose a ranking-based consistency constraint that greatly improves the model's optimization performance.
Moreover, we employ an orthogonality constraint instead of distance-based constraints to directly achieve linear independence between the global and personal item embedding tables.
Notably, our proposed Fed3CR is a plug-and-play method, which can be integrated with other FedRec methods to improve its performance.
Extensive experiments on four real-world datasets represent that Fed3CR improves the utilization efficiency of consensus and achieves superior performance.

The contributions of Fed3CR are as follows:
\begin{itemize}

\item We propose Fed3CR, which mitigates FedRec's statistical heterogeneity problem by increasing clients' efficiency in using consensus. It can be integrated with other FedRec methods to improve their performance.

\item We are the first to study the consensus degradation problem in FedRec with mathematical analysis and discussion. Based on this, we propose an ACE strategy to learn the relationship between global and personal item embeddings, adaptively enhancing consensus for clients.

\item We are the first to systematically promote consistency and complementarity between global and personal views in FedRec, effectively improving fusion efficiency.

\item Extensive experiments show that Fed3CR performs competitively against the baseline,  significantly improving the client's ability of utilizing consensus.
   
\end{itemize}

\section{Related Work}

\subsection{Multi-view Learning for Recommendation}

Introducing multi-view learning into the research of recommendation systems has shown significant potential \cite{xiao2019knowledge, ma2024multicbr}.
In real-world scenarios, user decisions are often influenced by both group behavioral patterns and individual unique preferences.
For example, after watching specific videos (\textit{e.g.}, "Harry Potter" movies), a considerable proportion of user groups tend to read related content (\textit{e.g.}, the original novels), which reveals the group correlation of user behavior.
Besides, there are often potential feature correlations between items in different fields or modalities, such as the core themes and worldviews shared by the "Harry Potter" movies and their book versions.
Therefore, many researchers integrate this heterogeneous information  of users and items using multi-view learning, effectively improving the representation of user and item embeddings, and significantly alleviating the data sparsity and cold start challenges of recommender systems.

Some research efforts target the cross-domain recommendation field, separating and learning the common and unique features of different domains \cite{gdccgr, TKDE_cdr1, TOIS_cdr2, TKDE_cdr3, TKDE_cdr4, TKDE_cdr5}. 
For example, CMVCDR \cite{TOIS_cdr2} respectively utilizes a static view and a sequential view to model users' general interests and current interests, and represents users' general interests as domain-shared and domain-specific parts.
The authors learn better user representations by refining cross-domain knowledge related to domain-shared factors and reducing the noise of irrelevant factors.
CDRVAE \cite{TKDE_cdr1} designs a hybrid architecture VAE for cross-domain recommendations, simultaneously performing inter-domain and intra-domain modeling to learn the optimal user preference distribution.
Some work focuses on multimodal recommendation, naturally treating different modal features of items as different views.
For example, MV-RNN \cite{TKDE_mmrec1} proposed a multi-view recurrent neural network, which integrates multimodal features to alleviate the cold start problem of items and effectively captures users' dynamic interests.
LGMRec \cite{lgmrec} proposed a multimodal recommendation framework guided by local and global graphs, jointly learning user behavior from personal and global perspectives, realizing robust interest modeling.
In addition, BayMAN \cite{BAYMAN} effectively combines the semantic and distance features by modeling different POI views, achieving more comprehensive user modeling and more robust recommendations.
Although these approaches demonstrate superior performance in different scenarios of recommender systems, they need to collect large-scale user-related data, violating the privacy requirement.
How to effectively utilize multi-view features under the premise of privacy protection still needs further exploration.

\subsection{Personalized Federated Recommendation} 
To avoid collecting user data while personalizing recommendations, FCF\cite{FCF}  introduces the FL framework to collaborative filtering techniques.
By maintaining the user embedding table locally to keep personalization, while the item embedding table is trained distributedly among servers and clients, FCF achieves privacy protection at the user level.
Then, many researchers explore based on this distributed paradigm, aiming to improve  communication efficiency, privacy protection capabilities and client incentives in personalized federated recommendations.
These approaches improve the performance of personalized federated recommendations in several dimensions, promoting the security and universality of the field. 
However, how to align client-private user representations and system-shared item representations for more accurate recommendations remains a research priority.
Specifically, due to the significant statistical heterogeneity among clients in FedRec, clients need to optimize local models to adapt to their unique interaction preferences.
This situation poses a challenge for server aggregation: the conflict of project embedding gradients uploaded by clients leads to performance drops.

To address these issues, some methods \cite{intro4_fedfast, perfedrec,co, fl2,fl3, preffedpoi} examine the clients from multiple views, cluster them based on relevance, and design cluster-level aggregation methods to mitigate statistical heterogeneity challenges.
For example, PerFedRec \cite{perfedrec} constructs a collaborative graph, utilizing federated GNN technology and attribute information  to learn the preferences of client groups for better personalization jointly.
PrefFedPOI \cite{preffedpoi} designs a reinforcement learning-based clustering mechanism for POI recommendation, which encourages active knowledge sharing by capturing fine-grained preference correlations among clients. It effectively improves the POI recommendation performance in data-sparse scenarios.
However, due to the need to collect partial attribute data from clients, even though these methods can still effectively protect clients' real interactions, they still cause some privacy concerns (\textit{e.g.}, attribute inference attacks \cite{zhang2023comprehensive}).

\subsection{Multi-view Federated Recommendation} 

There is an intuitive coupling between multi-view learning and federated recommendation due to the ability to separate different views of item embeddings.
Specifically, examining item embeddings from both global and local views can effectively capture the common features of items and the most attractive features to clients.
Compared to clustering-based methods, these methods essentially use learnable intermediate variables as a bridge to connect uniform item embeddings and unique client embeddings, thereby effectively mitigating their statistical heterogeneity.

FedDCSR \cite{zhang2024feddcsr} and FedHCDR \cite{zhang2024fedhcdr} target cross-domain federated recommendation tasks,  designing methods to learn domain-shared and domain-specific features of items, enabling efficient cross-domain alignment and sufficient personalization.
Some approaches \cite{fl1, pfedrec, gpfedrec, fedrap} focus on personalized federated recommendation tasks and design methods to achieve user-level personalization. 
For example, PFedRec \cite{pfedrec} introduced a dual personalization technique, generating local views of global item embeddings for each client, enabling fine-grained personalization.
To better understand the relationships between clients, GPFedRec \cite{gpfedrec} constructs a global user relationship graph while maintaining privacy, realizing graph-guided personalized item embedding fine-tuning, which effectively improves the system's recommendation capability.
FedRAP \cite{fedrap} separates the global-shared and client-specific views of item embeddings, encouraging divergence between them to obtain better representations.
Despite the advancements of these methods, their utility is limited by a lack of systematic consideration for the efficient utilization of global item embeddings and the relationship between the two views.
To this end, we propose Fed3CR to adaptively enhance the client-specific global information to improve the utilization efficiency of consensus for each client.
It also provides better decoupling by encouraging the two views to be semantically spatially consistent while acquiring different features, effectively improving the representations after the fusion.

\section{Degraded Consensus}

\begin{figure}[t]
	\centering
	\includegraphics[width=0.95\linewidth]{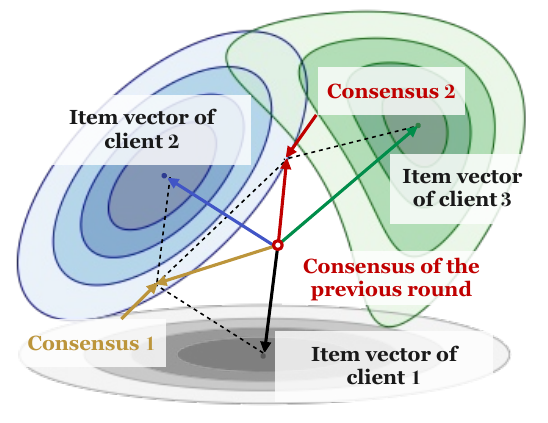}
	\caption{A toy example of the aggregation process in the server, from which we can obtain the following observations: 1) The final aggregated consensus (the red arrow) degrades compared to any single vector (the blue, green, black ones). 2) It is more challenging for client 1 to extract useful information from the consensus compared to clients 2 and 3. This observation shows that the degradation level of the consensus varies for each client. 3) Comparing yellow and red arrows, the more clients involved in the aggregation, the more severe the consensus degrades.}
	\label{fig:toy}
\end{figure}

To illustrate the consensus degradation phenomenon in the server, we provide a toy example in Figure~\ref{fig:toy} that includes the item distribution uploaded by three clients.
Among them, black, blue, and green arrows represent the item embedding of the clients, the yellow arrow represents the consensus aggregated by clients 1 and 2, and the red one represents the consensus aggregated by all clients.
In FL architectures, the server utilizes the item embeddings uploaded by the clients to update the consensus formed in the previous round.
Intuitively, the final aggregated consensus (the red arrow) degrades compared to any single vector. 
Moreover, it is more challenging for client 1 to extract helpful information from the consensus than for clients 2 and 3. 
This observation shows that the degradation level of the consensus varies for each client.
Additionally, comparing the yellow and red arrows, the more clients involved in the aggregation, the more severe the consensus degrades.

Furthermore,  in order to probe deeper into the causes of consensus degradation, we make some theoretical clarifications.
First, we make the following assumption.
\begin{assumption}
\label{assumption1}
Suppose $\mathcal{C}_i$ denotes the global item embedding table updated in each client and $\mathcal{L}_i(\cdot)$ denotes the loss function. Each client $i$ satisfies the following assumptions:
\begin{enumerate}
    \item \textit{$\mathcal{L}_i(\mathcal{C}_i)$ is convex.}
    \item \textit{$\mathcal{L}_i(\mathcal{C}_i)$ is $\rho$-Lipschitz, which means $\left\| \mathcal{L}_i(\mathcal{C}_i) - \mathcal{L}_i(\mathcal{C}_i')\right\| \le \rho \left\| \mathcal{C}_i - \mathcal{C}_i'\right\|$} for any $\mathcal{C}_i$, $\mathcal{C}_i'$.
    \item \textit{$\mathcal{L}_i(\mathcal{C}_i)$ is $\beta$-smooth, means $\left\| \nabla\mathcal{L}_i(\mathcal{C}_i) - \nabla\mathcal{L}_i(\mathcal{C}_i')\right\| \le \beta \left\| \mathcal{C}_i - \mathcal{C}_i'\right\|$ for any $\mathcal{C}_i$, $\mathcal{C}_i'$.}
\end{enumerate}
\end{assumption}

Assume $\mathcal{C}_i^*$ represents the optimal global item embedding table of the $i$-th client. The difference between the aggregated item embedding $\mathcal{C}$ and $\mathcal{C}_i^*$ can be formulated as:
  \begin{equation}\label{e-appendix1}
  \begin{aligned}
\left \| \mathcal{C}  - \mathcal{C}_i^* \right \|  &= \left \| \frac{1}{\left | \mathcal{U}_s  \right | } \sum_j \mathcal{C}_j^* - \mathcal{C}_i^*\right \|   = \left \| \frac{1}{\left | \mathcal{U}_s  \right | } \sum_j \left(\mathcal{C}_j^* - \mathcal{C}_i^*\right) \right \|  \\
&\le \frac{1}{\left | \mathcal{U}_s  \right | } \sum_j \left\| \mathcal{C}_j^* - \mathcal{C}_i^*\right \|,
  \end{aligned}
  \end{equation}
  where $\mathcal{U}_s$ represents the set of clients participating in this federated training round~\cite{fedavg}.
Since $\mathcal{C}_j^*$ is the local optimum of the $j$-th client, the gradient $\nabla\mathcal{L}_j(\mathcal{C}_j^*)=0$.
Based on the theory of Taylor expansion, we establish a connection between $\mathcal{C}_j^*$ and $\mathcal{C}_i^*$, which is denoted as:
  \begin{equation}\label{e-appendix2}
  \begin{aligned}
\nabla\mathcal{L}_j(\mathcal{C}_i^*) &= \nabla\mathcal{L}_j(\mathcal{C}_j^*) + \nabla^2\mathcal{L}_j(\xi)\left( \mathcal{C}_j^* - \mathcal{C}_i^*\right) \\
&= \nabla^2\mathcal{L}_j(\xi)\left( \mathcal{C}_j^* - \mathcal{C}_i^*\right),
  \end{aligned}
  \end{equation}
where $\xi$ is a solution between $\mathcal{C}_j^*$ and $\mathcal{C}_i^*$. $\nabla^2\mathcal{L}_j(\cdot)$ denotes the second order derivative matrix (Hessian matrix). 
According to assumption~\ref{assumption1}, Hessian matrix $\nabla^2\mathcal{L}_j(\cdot)$ is general positive semi-definite. To conveniently discuss the source of consensus degradation, we first assume that $\nabla^2\mathcal{L}_j(\cdot)$ is positive definite here. Thus, we have:
  \begin{equation}\label{e-appendix3}
  \begin{aligned}
\mathcal{C}_j^* - \mathcal{C}_i^* = \left(\nabla^2\mathcal{L}_j(\xi)\right)^{-1} \nabla\mathcal{L}_j(\mathcal{C}_i^*).
  \end{aligned}
  \end{equation}
  Since $\nabla^2\mathcal{L}_j(\cdot)$ is positive definite, we can obtain that $\left\|\left(\nabla^2\mathcal{L}_j(\xi)\right)^{-1}\right\| \le 1 / \lambda_{min}$, where $\lambda_{min}$ is the minimum eigenvalue of $\nabla^2\mathcal{L}_j(\xi)$. 

  Thus, we can obtain an upper bound of Eq.~(\ref{e-appendix1}), which can be denoted as:
    \begin{equation}\label{e-appendix4}
  \begin{aligned}
\left \| \mathcal{C}  - \mathcal{C}_i^* \right \|  \le \frac{1}{\lambda_{min}\left|\mathcal{U}_s\right|}\sum_j\left \|\nabla\mathcal{L}_j(\mathcal{C}_i^*) \right \| .
  \end{aligned}
  \end{equation}

  We denote the difference between the gradients between clients $i$ and $j$ as $\delta_{ij}$. Then, we have:
      \begin{equation}\label{e-appendix5}
  \begin{aligned}
\left \| \nabla\mathcal{L}_j(\mathcal{C}_i^*) \right \| \le \left \| \nabla\mathcal{L}_j\left( \mathcal{C}_i^* \right)  - \nabla\mathcal{L}_i\left( \mathcal{C}_i^* \right) \right \| + \left \| \nabla\mathcal{L}_i(\mathcal{C}_i^*) \right \|= \delta_{ij} .
  \end{aligned}
  \end{equation}
Substituting this into Eq.~(\ref{e-appendix4}), the upper bound can be formulated as:
      \begin{equation}\label{e-appendix6}
  \begin{aligned}
\left \| \mathcal{C}  - \mathcal{C}_i^* \right \|  \le \frac{1}{\lambda_{min}\left|\mathcal{U}_s\right|}\sum_j\delta_{ij} \le \delta_i,
  \end{aligned}
  \end{equation}
where $\delta_i=\frac{1}{\left|\mathcal{U}_s\right|}\sum_j\delta_{ij}$ denotes the average distribution difference between the $i$-th client and the others.

Then, considering a more general situation where $\nabla^2\mathcal{L}_j(\cdot)$ is positive semi-definite. We can still find a $\xi$ for which Hessian matrix is positive definite in a small neighbourhood of $\xi$. Therefore, the above conclusion is still valid.

In summary, the consensus after server aggregation differs from the optimal global item embeddings of any client. The magnitude of this difference is determined by the average distribution difference between this client and others, the larger difference is, the harder it is for consensus to fit into this client distribution.
However, client interactions in personalized FedRec vary, making it challenging to adapt the consensus to any client, \textit{i.e.}, the consensus degrades.
Since the degradation of consensus is personalized, it is difficult to learn a uniform consensus for all clients. Therefore, it is necessary to adaptively perform consensus enhancement based on the degree of consensus degradation for each client.

\begin{figure*}[t]
\centering
\includegraphics[width=1\textwidth]{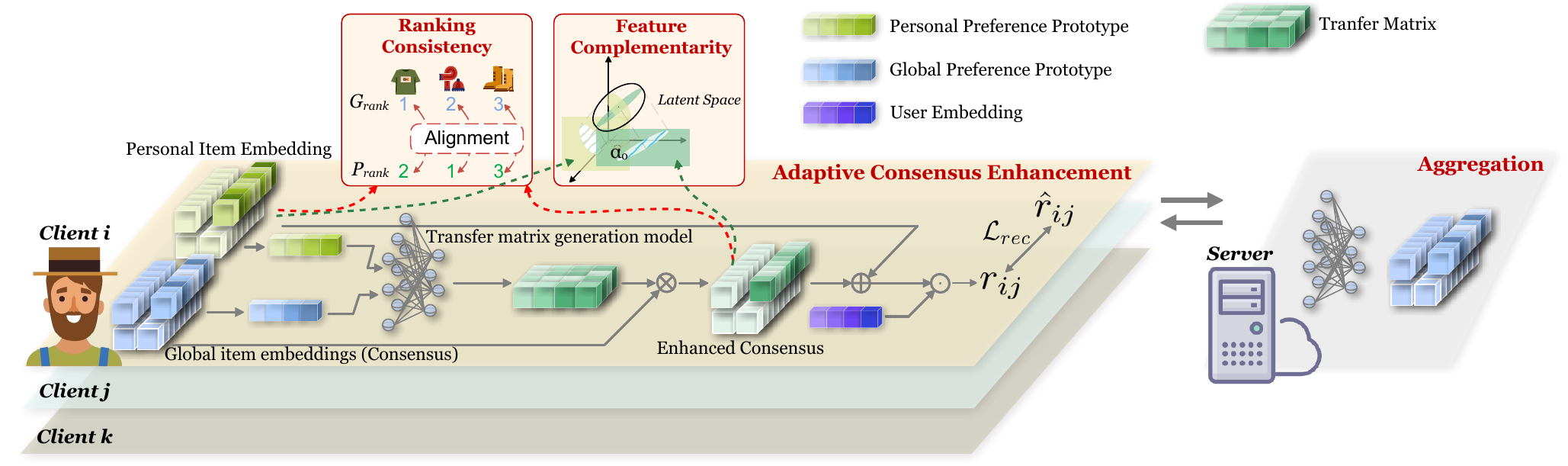}
\vspace{-0.1cm}
\caption{Overview of our proposed Fed3CR architecture, in which  \textit{$G_{rank}$} and \textit{$P_{rank}$ }denotes the client's preference rank in global and personal views.}
\vspace{-0.3cm}
\label{fig:2}
\end{figure*}
\section{Methodology}
\subsection{Preliminary}

Consider a typical personalized FedRec framework comprising a server and $N$ clients, where each client denotes a single user with personal interactions.
Assume that $ \mathcal{C}=\{\textbf{c}_{j}\}_{j=1}^{M}$ represents the global item embedding table (\textit{i.e.},~consensus) aggregated in the server.
The client $i$ in Fed3CR holds four types of parameters (\textit{i.e.}, user embedding $\textbf{u}_i$, global item embedding table $ \mathcal{C}_i$, personal item embedding table $\mathcal{V}_i$ and parameters $\theta_i$ in transfer matrix generation model $\mathcal{F}_i(\theta_i )$) and personal interaction matrix $\mathcal{R}_i \in \mathbb{R} ^{N\times M}$. 
All the embeddings are with dimension $d$. 
Each client trains them with personal interactions, sending $ \mathcal{C}_i$ and $\theta_i$ to the server for aggregation and collaborative training.
After aggregation, the server randomly selects a subset $\mathcal{U}_s$ of clients, sending them the updated $\mathcal{C}$ to improve training efficiency~\cite{fedavg}.
As a federated learning process, we aim to minimize the loss between the predictions and the ground truths for the $i$-th client, which is formulated as:
\begin{equation}\label{e-1} 
  \mathcal{L}_{rec}= -\sum_{(i, j)\in \mathcal{R}_i}r_{ij}log\hat{r}_{ij}+(1-r_{ij})log(1-\hat{r}_{ij}),
\end{equation}
where $(i, j)$ is a user-item pair in $\mathcal{R}_i$.
We focus on the more common and basic implicit recommendation tasks, where user interactions with projects are typically behaviors such as clicks, views, purchases, \textit{etc}.
Therefore, $\mathcal{R}_i$ is a binary matrix, where $r_{ij}=1$ denotes the interaction occurs and $r_{ij}=0$ vice versa. $\hat{r}_{ij}$ represents the predicted probability of client $i$ interacting with item $j$. 

\subsection{Overview of Fed3CR}

The proposed Fed3CR architecture is shown in Figure~\ref{fig:2}. Clients in Fed3CR learn both global and personal item embedding tables, fusing them for personalized recommendations.
To improve the effectiveness of consensus, we view the server as a teacher who imparts general consensus without personalized instruction, and introduce a module named \textbf{A}daptive \textbf{C}onsensus \textbf{E}nhancement (ACE) for each client.
In this module, we first look up the global and personal item embedding tables based on interactions to generate global and personal prototypes of preference.
Then, the client uses the transfer matrix generation model to learn the relationship between the two prototypes and generates a transfer matrix to enhance the consensus to fit its personal distribution adaptively.
Furthermore, to improve fusion efficiency and fully decouple global and personal item embeddings, we employ a \textbf{C}onsistency- and \textbf{C}omplementarity-aware \textbf{O}ptimization (C2O) strategy.
It comprehensively improves both consistency and complementarity between the two views.
After local training, only the global item embedding table and the transfer matrix generation model parameters are communicated between the server and clients for better personalization. The pseudo-code of Fed3CR is shown in Algorithm~\ref{alg-1}.

\begin{algorithm}[t]
\caption{Fed3CR}\label{alg-1}
\begin{flushleft}
    \textbf{Server Procedure}:
\end{flushleft}
\begin{algorithmic}[1]
    \State Initialize global item embedding table $\mathcal{C}$
    \For{$t=1,2,\dots, T$} \textcolor{blue}{\Comment{Federated communication rounds}}
        \State Randomly select a subset $\mathcal{U}_s$ of clients
        \For{client $i \in \mathcal{U}_s$  \textbf{in parallel}}
            \State $\mathcal{C}_i, \theta_i \gets \textbf{LocalUpdate}(\mathcal{C})$ 
        \EndFor
        \State $\mathcal{C}= \frac{1}{\left | \mathcal{U}_s  \right | } \sum_{i\in \mathcal{U}_s} \mathcal{C}_i$ by Eq.~(\ref{e-2})\textcolor{blue}{\Comment{Aggregate consensus}}
        \State $\theta= \frac{1}{\left | \mathcal{U}_s  \right | } \sum_{i\in \mathcal{U}_s} \theta_i$ \textcolor{blue}{\Comment{Aggregate parameters}}
    \EndFor
    \Statex
\end{algorithmic}
\begin{flushleft}
    \textbf{LocalUpdate$(\mathcal{C})$:}
\end{flushleft}
\begin{algorithmic}[1]
    \State $\mathcal{C}_i \leftarrow \mathcal{C}$ \textcolor{blue}{\Comment{Update global item embedding table}}
    \State $\theta_i \leftarrow \theta$ \textcolor{blue}{\Comment{Update parameters of $\mathcal{F}_{i}\left( \cdot  \right)$}}
    \For{$e=1,2,\dots,E$} \textcolor{blue}{\Comment{Client's training iteration}}
        \State Compute $\textbf{p}_{i}^G$ and $\textbf{p}_{i}^P$ by Eq. (\ref{e-4})

        \State Generate personalized tranfer matrix $\mathbf{ W}_{i}$ by Eq. (\ref{e-3})
        \State Compute $\mathcal{C}^E_i$ by Eq. (\ref{e-5})
        \State Compute $\mathcal{L}_a$ via Eq. (\ref{e-13}) \textcolor{blue}{\Comment{For alignment}}
        \State Compute $\mathcal{L}_o$ via Eq. (\ref{e-15})  \textcolor{blue}{\Comment{For orthogonality}}
        \State $\mathcal{V}_{i}^F= \mathcal{C}_{i}^E + \mathcal{V}_{i}$ by Eq. (\ref{e-6})
        \State Computer $\hat{r}_{ij}$ for each candidate item $\textbf{v}_{ij}$ via Eq. (\ref{e-rs}) 
        \State Compute $\mathcal{L}_{rec}$ via Eq. (\ref{e-1})\textcolor{blue}{\Comment{For recommendation}}
        \State Update $\mathcal{C}_i$, $\mathcal{V}_i$ and parameters $ \theta_i$ of $\mathcal{F}_{i}\left( \cdot  \right)$
    \EndFor

    \State \textbf{Return} $\mathcal{C}_i$, $\theta_i$
\end{algorithmic}
\end{algorithm}

\subsection{Adaptive Consensus Enhancement}
We start from the consensus aggregation process on the server, which can be formulated as:
\begin{equation}\label{e-2} 
  \mathcal{C}= \frac{1}{\left | \mathcal{U}_s  \right | } \sum_{i\in \mathcal{U}_s} \mathcal{C}_i,
\end{equation}
where $\mathcal{C}_i$ denotes the global item embedding table updated and uploaded by the $i$-th client.
As mentioned above, statistical heterogeneity challenge among clients leads to degraded consensus in servers, 
Since the issue cannot be eliminated entirely,  we consider enhancing the consensus on the client side to fit its distribution better.
However, learning how to enhance consensus stably and efficiently for a specific client is non-trivial since the server randomly selects clients to participate in training each communication round (between the server and the clients).
Hence, given the semantic consistency of client preferences in global and personal views, we make the client learn how to adaptively enhance the consensus itself, which is based on the insight of~\cite{ptupcdr}. 

Specifically, we employ a transfer matrix generation model to learn the relationship between the client's global and personal preference prototypes, adaptively generating a transfer matrix. This process is formulated as:
\begin{equation}\label{e-3} 
\mathbf{ W}_{i}= \mathcal{F}_{i}\left( \left[\textbf{p}_{i}^G, \textbf{p}_{i}^P\right] |\theta_i  \right),
  \end{equation}
where $\mathbf{ W}_i \in \mathbb{R}^{d\times d} $ denotes the transfer matrix generated by the transfer matrix generation model $\mathcal{F}_i(\theta_i)$.
$\textbf{p}_{i}^G$ and $\textbf{p}_{i}^P$ denotes the client's preference prototype in the global and personal view, respectively, which is formulated as:
  \begin{equation}\label{e-4}
  \begin{aligned}
 \textbf{p}_{i}^G&=\frac{1}{\left |\mathcal{C}_i^{\left(+\right)}\right|}\sum_{\textbf{c}_{ij} \in \mathcal{C}_i^{\left(+\right)} }\textbf{c}_{ij},\\
\textbf{p}_{i}^P&=\frac{1}{\left |\mathcal{V}_i^{\left(+\right)}\right|}\sum_{\textbf{v}_{ij} \in \mathcal{V}_{i}^{\left(+\right)} }\textbf{v}_{ij},     
  \end{aligned}
  \end{equation}
where $\mathcal{C}_i^{\left(+\right)}$ and $\mathcal{V}_i^{\left(+\right)}$ denotes the global and personal representation of interacted items on the client, respectively. And $\left[\textbf{p}_{i}^G, \textbf{p}_{i}^P\right]$ in Eq.~(\ref{e-3}) represents the concatenate operation of prototypes.
By collaboratively learning these preference prototypes, $\mathcal{F}_i(\theta_i)$ can efficiently generate a transfer matrix $\mathbf{ W}_i$, even in the presence of significant uncertainty caused by the client selection process.

 Subsequently, we can calculate the client-specific enhanced global item embedding table $\mathcal{C}^E_{i}$, which is formulated as:
  \begin{equation}\label{e-5} 
\mathcal{C}^E_{i}= \mathbf{ W}_{i}\mathcal{C}_{i}.
  \end{equation}
Clearly, $\mathcal{C}^E_{i}$ denotes the enhanced consensus.
Compared to $\mathcal{C}_{i}$, $\mathcal{C}^E_{i}$ represents client $i$'s unique view on the consensus and is more suitable for its distribution. The ACE strategy effectively improves the utilization of $\mathcal{C}$, which helps the client achieve better personalization, even under sparse interactions and distribution uncertainties.

Finally, we add $\mathcal{C}_i^E$ to $\mathcal{V}_i$ to obtain the fused item embedding table $\mathcal{V}_i$ in the client $i$, which can be formulated as:
  \begin{equation}\label{e-6} 
\mathcal{V}_{i}^F= \mathcal{C}_{i}^E + \mathcal{V}_{i}.
  \end{equation}
As a result, $\mathcal{V}_{i}^F$ contains the client's unique perspective of consensus and its individual preferences of items. The predicted interaction probability for the $j$-th item is calculated by:
  \begin{equation}\label{e-rs} 
\hat{r}_{ij}=\sigma(\textbf{u}_i^T\textbf{v}_j),
  \end{equation}
  where $\textbf{v}_j$ is the vector in $\mathcal{V}_{i}^F$, $\sigma(\cdot)$ is the Sigmoid operation.
Compared with other approaches, the core insight of the ACE module is to adaptively generate optimal consensus for each client.
This mechanism makes the more effective representation of consensus, rather than just a simple average of individual local models, thus improving the efficiency of consensus utilization.

\subsection{Consistency- and Complementarity-aware Optimization }
As discussed above, maintaining consistency and complementarity between the two views helps avoid introducing misinformation while improving the fusion efficiency. We now detail the proposed consistency- and complementarity-aware optimization strategies.

\subsubsection{Consistency}
An intuitive approach to promote consistency is to ensure the similarity (\textit{i.e.}, cosine similarity) between any two vectors in $\mathcal{C}_i^E$ is equal to their similarity in $\mathcal{V}_i$, which is denoted as:
  \begin{equation}\label{e-11} 
\frac{\mathcal{C}^E_i\left (\mathcal{C}^E_i\right )^T}{\left \| \mathcal{C}^E_i \right \| ^{2} } =\frac{\mathcal{V}_i\left (\mathcal{V}_i\right )^T}{\left \| \mathcal{V}_i \right \| ^{2} } .
  \end{equation}
However, finding an optimal solution for Eq.~(\ref{e-11}) is exceedingly difficult. First, in personalized FedRec, the global and personal distributions differ significantly, making it complex to recommend while preserving similarities, especially in the early stages of training. Second, the non-convexity of the cosine similarity function makes the model prone to converging to local optima, leading to a suboptimal solution. To this end, inspired by the insights of ListNet~\cite{listnet}, we focus on keeping the relative ranking relationships of vectors between vectors and corresponding prototypes in $\mathcal{C}_i^E$ and $\mathcal{V}_i$.

 Taking $\textbf{p}_i^P$ and $\mathcal{V}_i$ as an example. We compute the top one probability for each pair $\left( \textbf{p}_i^P, \textbf{v}_{ij}\right
 )$, $\textbf{v}_{ij} \in \mathcal{V}_i$. It represents the probability that $j$-th item is most likely to be interacted with by the client $i$. This process can be formulated as:
   \begin{equation}\label{e-12}
  \begin{aligned}
 \mathcal{P}\left( \textbf{p}_i^P, \textbf{v}_{ij}\right
 )=  \frac{\text{CosineSimilarity}\left(\textbf{p}_i^P, \textbf{v}_{ij} \right)}{\sum_{k=1}^M \text{CosineSimilarity}\left(\textbf{p}_i^P, \textbf{v}_{ik} \right)}.    
  \end{aligned}
  \end{equation}
 Similarly, we can compute the top one probability for each $\left( \textbf{p}_i^E, \textbf{c}_{ij}^E\right
 )$ as $ \mathcal{P}\left( \textbf{p}_i^E, \textbf{c}_{ij}^E\right
 )$, where $\textbf{p}_i^E$ represents the global preference representation after being mapped by $\mathbf{ W}_i$, and $\textbf{c}_{ij}^E \in \mathcal{C}_i^E$.
 Our objective is to encourage $\mathcal{P}\left( \textbf{p}_i^E, \textbf{c}_{ij}^E\right
 )$ and $\mathcal{P}\left( \textbf{p}_i^P, \textbf{v}_{ij}\right
 )$ to be as close as possible, thus promoting consistency between $\mathcal{C}_i^E$ and $\mathcal{V}_i$.
Therefore, the consistency constraint can be formulated as:
  \begin{equation}\label{e-13} 
  \begin{aligned}
\mathcal{L}_{a}=&-\frac{1}{2} \sum_j \mathcal{P}\left( \textbf{p}_i^P, \textbf{v}_{ij}\right
 )\log{\mathcal{P}\left( \textbf{p}_i^E, \textbf{c}_{ij}^E\right
 )}\\
&-\frac{1}{2}\sum_j\mathcal{P}\left( \textbf{p}_i^E, \textbf{c}_{ij}^E\right
 )\log{\mathcal{P}\left( \textbf{p}_i^P, \textbf{v}_{ij}\right
 )}.
  \end{aligned}
  \end{equation}
Compared to Eq.~(\ref{e-11}), the ranking loss provides a softer constraint, which is more conducive to model optimization.

\subsubsection{Complementarity}
Same as common settings~\cite{orth1, orth2, orth3}, the feature space of $\mathcal{V}_i$ should be orthogonal to that of $\mathcal{C}_i^E$, which is formulated as:
\begin{equation}\label{e-14} 
\left (\mathcal{C}_i^E\right )^T\mathcal{V}_i=\mathcal{O},
  \end{equation}
where $\left (\mathcal{C}_i^E\right )^T\mathcal{V}_i$ represents the correlation matrix $\mathcal{E}$, and $\mathcal{O}$ denotes the zero matrix.
Hence, in order to achieving a more effective addition, the complementarity constraint is formulated as:
 \begin{equation}\label{e-15} 
\mathcal{L}_{o}= \frac{1}{d}\sum_i \sum_{j} \left \|\mathcal{E}_{ij}\right \|   ^{2} .
  \end{equation}
Compared to distance-based complementarity constraints (\textit{e.g.},~Euclidean distance), $\mathcal{L}_{o}$ directly achieve linear independence by enforcing mutual orthogonality between features. Thus, it can promote $\mathcal{C}_i^E$ and $\mathcal{V}_i$ to capture different features of the items.

\subsubsection{Overall constraint}
  Overall, the optimization of Fed3CR framework is to minimize:
   \begin{equation}\label{e-16} 
\mathcal{L}_{Fed3CR}=\mathcal{L}_{rec} + \beta_a\mathcal{L}_{a}+\beta_o\mathcal{L}_{o},
  \end{equation}
  where $\beta_a$ and $\beta_o$ are hyperparameters. In this way, Fed3CR effectively decouple and align global and personal item embeddings, achieving better personalization.

\section{Experiments}

We conduct a comprehensive evaluation of the Fed3CR architecture for the following issues:
\begin{itemize}
    \item \textbf{\textit{RQ1} :} How does Fed3CR perform compared with other state-of-the-art methods?
    \item \textbf{\textit{RQ2} :} How do the key components of Fed3CR contribute to the performance?
    \item \textbf{\textit{RQ3} :}  How can Fed3CR maintain consistency and complementarity between global and personal item embeddings?
    \item \textbf{\textit{RQ4} :} Could the proposed Fed3CR mechanism improve the performance of other personalized FedRec methods?
    \item \textbf{\textit{RQ5} :} What are the impacts of the key hyperparameters on the performance of Fed3CR?
    
\end{itemize}

\subsection{Experiment Preparation}

\subsubsection{Datasets}
We validate the performance of Fed3CR on four real-world datasets, \textit{i.e.}, MovieLens-1M\footnote{\url{https://grouplens.org/datasets/movielens/}\label{dt_ml}}\cite{movielens}, LastFM-2K\footnote{\url{https://grouplens.org/datasets/hetrec-2011/}\label{dt_ml}} \cite{LastFM}, Douban-Book\footnote{\url{https://github.com/fengzhu1/GA-DTCDR/tree/main/Data/}\label{dt_ml}}~\cite{douban} and QB-article\footnote{\url{https://tenrec0.github.io}\label{dt_ml}} \cite{tenrec}. 
These datasets cover four world-wide recommendation domains: movies, music, books and articles, which consist of real user interactions and ratings.
Specifically, we use LastFM-2K and Douban-Book datasets to quickly validate the proposed method. 
We evaluate the performances of the methods in the multi-client with high sparsity scenario using the QB-article dataset and verify the methods performances in the low interaction sparsity environment through the MovieLens-1M dataset.
Following~\cite{fedrap}, we filter out users with interactions less than 5 from Douban-Book, 10 from LastFM-2K, MovieLens-1M 
and QB-article. 
 The statistics of four datasets are shown in Table~\ref{t-dataset}.

\subsubsection{Baselines}
To illustrate the upper bound of performance for FedRec methods, we compare Fed3CR with the most typical \textbf{Centralized methods}: NeuMF \cite{neumf} and LightGCN \cite{lightgcn}. 
To illustrate the  performance for Fed3CR, we compare it with the following \textbf{Federated methods}: FedMF\cite{fedmf}, FedNCF\cite{fedncf}, PFedRec\cite{pfedrec}, PerFedRec\cite{perfedrec}, GPFedRec\cite{gpfedrec} and FedRAP\cite{fedrap}.

The details of the baselines are introduced as follows:
\textbf{Centralized methods}
\begin{itemize}
    \item \textbf{NeuMF}~\cite{neumf}, the most typical DNN-based recommendation method, specializes in learning non-linear relationships between users and items.
    \item \textbf{LightGCN}~\cite{lightgcn}, the most typical GNN-based recommendation approach, which exploits the higher-order relationships on the user-item interaction graph.
\end{itemize}
\textbf{Federated methods}
\begin{itemize}
    \item \textbf{FedMF}~\cite{fedmf}, which introduces matrix factorization to FL architecture, allowing users to upload only homomorphic encrypted gradients.
    \item \textbf{FedNCF}~\cite{fedncf}, which extends NeuMF to FL architecture to capture nonlinear user-item interactions in a privacy-preserving way. 
    \item \textbf{PFedRec}~\cite{pfedrec}, which proposes a dual personalization mechanism, achieving personalization of both the model and embedding tables within each client.
    \item \textbf{PerFedRec}~\cite{perfedrec}, which presents a GNN-based personalized FedRec method, achieving hierarchical personalization at global, cluster, and its own levels for each client through collaborative graphs.
    \item \textbf{GPFedRec}~\cite{gpfedrec}, which leverages a client-relation graph constructed on the server (without accessing its interaction records) to achieve graph-guided personalization.
    \item \textbf{FedRAP}~\cite{fedrap}, which enables additive personalization on the client side by decoupling global and personal item embeddings.
\end{itemize}

\subsubsection{Evaluation Metrics}  We use two widely used evaluation metrics for Top-K recommendation: HR@K (Hit Ratio) and NDCG@K (Normalized Discounted Cumulative Gain \cite{ndcg}).  HR@K indicates the accuracy of the Top-K recommendation, while NDCG@K demonstrates the quality of the Top-K ranking. We set $K=10$ in the following experiments.
Additionally, we use Top-K' Rank Biased Overlap (RBO) \cite{RBO} to verify whether the proposed method can improve the semantic-level similarity between the two views in the client. RBO simultaneously measures the overlap and ranking of the Top-K' item lists for personal and global views, with a value between 0 and 1, where 1 indicates that the lists are completely identical and in the same order. For a comprehensive ebaluation, We set the hyperparameter of RBO as $p=0.99$.

\begin{table}[t]
\setlength{\abovecaptionskip}{0.2cm}
\caption{Statistic details of four datasets. The \#Avg. is the average number of interactions per user. Sparsity is percentage of \#Interactions in (\#Clients × \#Items)}
\centering
\scalebox{0.95}{
\begin{tabular}{lccccc}
\toprule 

\rowcolor{mygrey}
Dataset & \#Clients & \#Items & \#Interactions & \#Avg. & Sparsity \\
\midrule 
        
        LastFM-2K & 1874 & 17612 & 92,780 & 50 & 99.72\% \\
        QB-article & 11368 & 6538 & 266,356 & 23 & 99.64\% \\
        Douban-Book & 1696 & 6777 & 95,107 & 56 & 99.13\% \\
        MovieLens-1M & 6040 & 3706 & 1,000,209 & 166 & 95.53\% \\
\bottomrule

\end{tabular}}

\label{t-dataset}
\vspace{-0.3cm}
\end{table}

\subsubsection{Implementation Details}
All methods are implemented using PyTorch, and experiments are conducted on an RTX 4090 GPU. We use the original code released by the authors for a fair comparison.
For a fair comparison, we follow the experimental setups in FedRAP \cite{fedrap}, including communication and client training round, embedding dimension $d$, negatively sampling and dataset split strategies. Specifically, we set the maximum number $T$ of the server-client communications and the client-side training iterations $E$ as 100 and 10, separately.
We use the widely adopted leave-one-out strategy for dataset split, and randomly select 4 negative samples for each positive item for the training procedure. 
Besides, the transfer matrix generation model $\mathcal{F}_i(\theta_i)$ is set as a fully connected network. We set it with input dimensions $\left[2\times d, 4\times d\right]$, $\left[2\times d, 4\times d, 8\times d\right]$ and $\left[2\times d, 4\times d, 8\times d, 16\times d\right]$, and the output dimension of the last layer is $d\times d$.
 The length scale coefficient $\alpha$ in ACE is set to 10, And the hyperparameters $\beta_a$ and $\beta_o$ in C2O are searched in the range of \{0.1, 0.3, 0.5, 0.7, 1\}. We use SGD optimizer~\cite{SGD} and apply ExponentialLR scheduler~\cite{ee} during clients' learning schedules. 
  And all methods use fixed latent embedding dimensions of 32 and a batch size of 2048. 
For cross-validation, we repeat the experiments five times and reported the mean results.

\begin{table*}[t]
    \centering
    \caption{Performance comparison on four real-world datasets. \textit{Cen.} denotes the centralized baselines, while \textit{Fed.} denotes the federated baselines. Best in bold and second with underline. Due to the use of the same experimental setup and dataset, we reuse some results provided in FedRAP. 
    See details in Sec. \ref{sec-topk}.} 
   
    \scalebox{1}{
    \begin{tabular}{l|l||cc|cc|cc|cc|cc}
        \toprule
        \rowcolor{mygrey}
         \multicolumn{2}{c||}{} & \multicolumn{2}{c|}{\textbf{MovieLens-1M}} & \multicolumn{2}{c|}{\textbf{Lastfm-2K}} & \multicolumn{2}{c|}{\textbf{Douban-Book}} & \multicolumn{2}{c|}{\textbf{QB-article}} & \multicolumn{2}{c}{\textbf{Average}}\\
        \rowcolor{mygrey}
        \multicolumn{2}{c||}{\multirow{-2}{*}{\textbf{Methods}}}  & H@10 & N@10 & H@10 & N@10 & H@10 & N@10 & H@10 & N@10 & H@10 & N@10\\
        \midrule
        \midrule
        \multirow{2}{*}{\textbf{Cen.}} & \textbf{NeuMF} & 62.52 & 35.60 & 20.81 & 11.36 & 46.82& 24.43& 40.94& 19.61 & 42.77 & 22.75\\
        & \textbf{LightGCN} & 99.43  & 90.43 & 28.56 & 14.32 & 58.59& 40.66 & 61.09 & 31.85 & 61.92 & 44.32\\
        \midrule
        \multirow{6}{*}{\textbf{Fed.}} & \textbf{PFedRec}  &     {68.59} &  {40.19} &  {21.61} & {10.88} & 45.12 & 23.73 & {46.00} & {22.28} &  45.33& 24.27\\
        & \textbf{PerFedRec} &  60.86 & 45.62 & 19.87 & 8.91 & 36.82 & 19.23 & 37.15 & 18.65 & 38.68 & 23.10\\
        & \textbf{GPFedRec} &     {72.17} &  {43.61} & {20.78} &  {10.50} & 43.64 & 23.36 & 43.86 & 20.83 & 45.11 & 24.58\\  
        & \textbf{FedMF} &   {65.95}  &  {38.77} &  {20.04} &  {9.87} & 39.54 & 21.41 & {40.54} & {17.76} & 41.52 & 21.95\\
        & \textbf{FedNCF} &     {61.31} &  {34.94}  &  {20.14} &  {9.99} & 39.79 & 20.23 &  {40.52} & {19.31} & 40.44 & 21.12 \\

        & \textbf{FedRAP} &     \underline{93.24} &  \underline{71.87}  &  \underline{23.29} &  \underline{10.99} &  \underline{49.59} & \underline{29.14} & \underline{53.98} & \underline{24.75} & \underline{55.03} & \underline{34.19} \\
        & \textbf{Fed3CR} & \textbf{97.13}& \textbf{86.99}& \textbf{28.68} & \textbf{13.01}& \textbf{55.98}& \textbf{38.77} & \textbf{57.23} & \textbf{26.84} & \textbf{59.76} & \textbf{41.40}\\
        \midrule
        \rowcolor{myred}
        \multicolumn{2}{c||}{\textbf{\%Improvement}}  &  4.17\%$\uparrow$& 21.04\%$\uparrow$& 23.14\%$\uparrow$ & 18.38\%$\uparrow$ & 12.89\%$\uparrow$ & 33.05\%$\uparrow$ & 6.02\%$\uparrow$ & 8.44\%$\uparrow$ & 8.60\%$\uparrow$ & 21.09\%$\uparrow$\\
         \bottomrule
    \end{tabular}
    }
    
    \label{tab:my_label}

\end{table*}

\subsection{Top-K Recommendation Performance (\textbf{\textit{RQ1}})}
\label{sec-topk}
We compare the Top-10 recommendation performance of Fed3CR with other baselines on four real-world datasets Table \ref{tab:my_label}, leading to the following observations. 
1) Compared to centralized methods, FedRec methods suffer from the statistical heterogeneity issue and thus have poor recommendation performance.
Our proposed Fed3CR method is comparable to typical centralized recommendation methods on both the HR@10 and NDCG@10  metrics. It even surpasses them on some metrics, which is something that other FedRec methods cannot achieve.
This is because the centralized methods need to learn an average pattern to adapt to all clients, which sacrifices the precise fit to the needs of individuals when trying to generalize to the entire user group.
In contrast, the essence of FedRec is to train recommenders on client data. This allows local models to capture the unique preferences and behavioral patterns of each client (or group of users) at a finer level.
Our proposed Fed3CR further amplifies this localization advantage by improving clients' efficiency in utilizing consensus.
Additionally, Fed3CR is designed to better leverage the data heterogeneity among clients.
It allows local models to maintain a certain degree of diversity, while refining consensus through aggregation mechanisms.
Hence, the final recommender is more effective, especially in scenarios where heterogeneity features are particularly critical.
2) Consensus-based methods (\textit{i.e.}, PFedRec, GPFedRec, FedRAP, and Fed3CR) consistently outperform other FedRec baselines across all datasets.
It denotes that decoupling consensus and personalized item embedding in the client in FedRec can help mitigate the feature drift issue and improve the client's representational capabilities.
Compared to other FedRec methods, Fed3CR adaptively performs client-specific consensus enhancement, improving the utilization efficiency of global item embeddings in the client.
Moreover, it promotes the consistency and complementarity of global and personal item embedding tables, effectively improving the fused embeddings' representation quality.
3) The performance improvement of Fed3CR varies on different datasets.
On the MovieLens-1M dataset, Fed3CR has an absolute performance advantage, indicating that Fed3CR is able to learn effective client and item representations in a relatively dense dataset.
Meanwhile, the relative improvement of NDCG is greater, indicating that Fed3CR not only finds items that users prefer (HR@10), but also better sorts these items (NDCG@10).
Moreover, compared to the MovieLens-1M dataset, Lastfm-2K has sparser interactions, indicating a more challenging recommendation task.
The significant advantage of Fed3CR over other federated models on this sparser dataset is even more pronounced, highlighting the superior design in dealing with data sparsity.
\begin{table}[t]
    \centering
    \caption{Ablation studies on LastFM-2K and Douban-Book datasets. H@10 and N@10 indecate HR@10 and NDCG@10, respectively. See details in Sec. \ref{sec-ab}.}
    \scalebox{0.99}{
    \begin{tabular}{c||ccc|cc|cc}
        \toprule

        \rowcolor{mygrey}
          &    & \multicolumn{2}{c|}{\textbf{C2O}}& \multicolumn{2}{c|}{\textbf{LastFM-2K}} & \multicolumn{2}{c}{\textbf{Douban-Book}}\\
        \rowcolor{mygrey}
        & \multirow{-2}{*}{\textbf{ACE}}& $\mathcal{L}_a$ & $\mathcal{L}_o$ & H@10 & N@10 & H@10 &N@10 \\
\midrule
         \midrule
        C0 &    &   & & 23.02 & 10.58 & 49.10 & 28.98 \\

        C1 &  \Checkmark  &   & &26.91 & 12.02 & 53.67 & 36.49\\

         C2 &    & \Checkmark & & 23.59 & 10.94 & 49.81 & 29.79 \\
         C3 &    &  &\Checkmark & 23.87 & 11.06 & 49.75 & 29.66\\
         C4 &    & \Checkmark  &\Checkmark & 24.21 & 11.39 & 50.36 & 30.20\\
         C5 &\Checkmark    & \Checkmark & & 27.09 & 12.25 & 54.16 & 36.81\\
         C6 &\Checkmark    &  &\Checkmark & 27.26 & 12.37 & 54.30 & 37.01\\
         Fed3CR &\Checkmark    & \Checkmark &\Checkmark & 28.68 & 13.01 & 55.98 & 38.77\\

        \bottomrule
    \end{tabular}}

    \label{tab:Ablation}
\end{table}
\begin{figure}[t]
    \centering
    \subfloat[{Consistency}\label{ab1.sub.1}]{
        \includegraphics[width=4.04cm,height = 3.03cm]{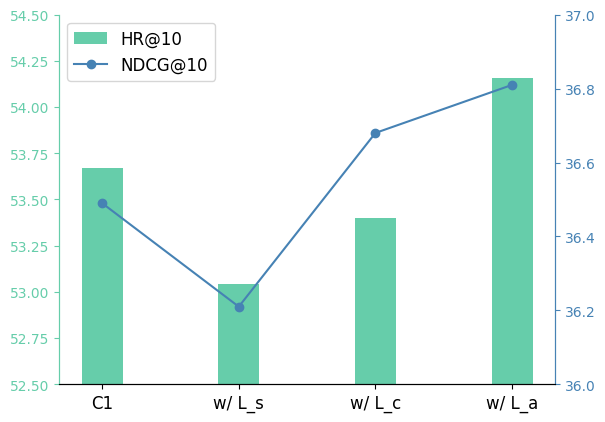}
        
        }
\subfloat[{Complementarity}\label{ab1.sub.2}]{
        \includegraphics[width=4.04cm,height = 3.03cm]{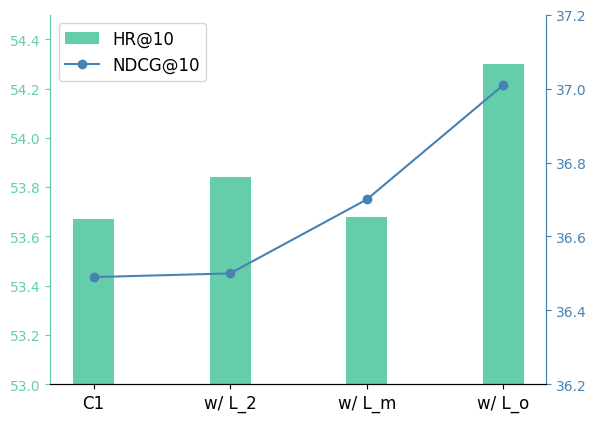}
        
        }

\caption{Ablation studies of consistency and complementarity constraints on the Douban-Book dataset, where C1 represents the variable in Table~\ref{tab:Ablation}. We replace the proposed consistency constraint (L\_a) with the commonly used similarity constraint (L\_s) and comparison (L\_c), and the complementarity constraint (L\_o) with the commonly used distance  (L\_2) and mutual information constraint (L\_m) for the experiments, respectively.  See details in Sec. \ref{sec-ab1}.}

\label{fig:AB1}
\end{figure}

\subsection{Ablation Studies of Key Components (\textbf{\textit{RQ2}})}
\subsubsection{Proposed module}
\label{sec-ab}
To demonstrate the performance of each component in Fed3CR, we conduct experiments on the LastFM-2K and Douban-Book datasets and presented the results in Table \ref{tab:Ablation}.
We use C0 to C6 to denote the ablative variants of Fed3CR. 
Accordingly, we obtain the following observations:
$\textcircled{1}$ The observed performance degradation upon removing any single module highlights the individual importance of each component.
The model's  performance drops significantly when the ACE module is not adopted.
This suggests that ACE is the most important component that significantly improves the fused item representation by improving the efficiency of the client's utilization of consensus.
$\textcircled{2}$ Both $\mathcal{L}_a$ and $\mathcal{L}_o$ in C2O module are effective, indicating that both the proposed consistency and complementarity constraints can improve the representational ability of the fused item embeddings.
Individually they limit effect, but jointly they outperform either constraint, suggesting their collaborative effect.
Combining ACE and C2O strategies together yields the best performance, demonstrating their synergistic contribution to recommendation performance. 
The above trends are generally consistent across the two evaluation metrics for both datasets, suggesting the  universality of the effectiveness of these components.

\begin{figure}[t]
    \centering
    \subfloat[{Consensus Transfer}\label{ab.sub.1}]{
        \includegraphics[width=3.5cm,height = 3cm]{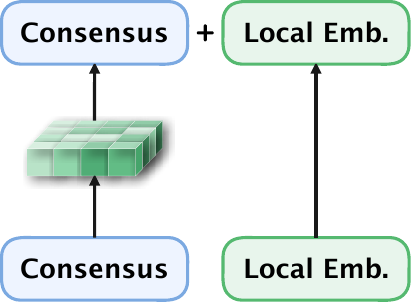}
        
        }
       \quad
    \subfloat[{Unified Transfer}\label{ab.sub.2}]{
        
        \includegraphics[width=3.5cm,height = 3cm]{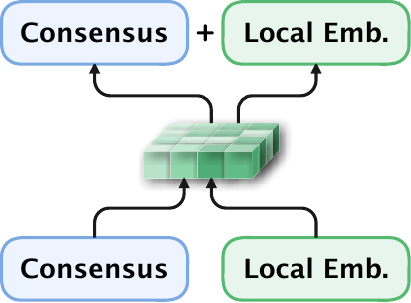}
    }

\caption{Commonly used consensus enhancement methods other than ACE module. The \textit{Local Emb.} denoted the personal item embeddings of each client. See details in Sec. \ref{sec-ab2}.}
\label{fig:AB2}
\end{figure}

\begin{table}[t]
    \centering
    \caption{Ablation studies of consensus enhancement methods on LastFM-2K and Douban-Book datasets. See details in Sec. \ref{sec-ab2}.}
    \scalebox{0.99}{
    \begin{tabular}{c||cc|cc}
        \toprule

        \rowcolor{mygrey}
            & \multicolumn{2}{c|}{\textbf{LastFM-2K}} & \multicolumn{2}{c}{\textbf{Douban-Book}}\\

        \rowcolor{mygrey}
        \multirow{-2}{*}{\textbf{Method}}& HR@10 & NDCG@10 & HR@10 &NDCG@10 \\
\midrule
         \midrule
        Consensus Transfer & 24.85 & 11.24 & 51.45 & 33.90\\
        Unified Transfer & 24.85 & 11.21 & 51.70 & 33.99\\
        ACE & 26.91 & 12.02 & 53.67 & 36.49\\

        \bottomrule
    \end{tabular}}

    \label{tab:Ablation2}
\end{table}

\subsubsection{Consistency and complementarity constraints}
\label{sec-ab1}

To demonstrate the performance of our proposed C2O module, we add the general consistency and complementarity constraints to the C1 variant mentioned in the previous section and conduct experiments on the Douban-Book dataset.
For consistency constraint, where L\_s and L\_c represent the general similarity and contrast constraints, and L\_a represents our strategies (variant C5 in Table~\ref{tab:Ablation}).
For complementarity constraint, where L\_2 and L\_m represent the L2 regularization of FedRec and mutual information constraints, and L\_o represents our strategies (variant C6 in Table~\ref{tab:Ablation}).
The results indicate that the proposed consistency and complementarity constraints outperform other constraints in both HR@10 and NDCG@10 metrics, proving the superiority of our C2O strategy.
Meanwhile, we observe that forcing consistency between consensus and personal item embedding leads to performance degradation, which is due to the fact that in the complex scenario of personalized FedRec, simply forcing complete consistency in understanding items between the two views leads to the loss of personalization.

\subsubsection{Consensus enhancement module}
\label{sec-ab2}

To further illustrate the performance of ACE module, we replace it with two commonly used consensus enhancement methods, as shown in Figure~\ref{fig:AB2}.
Among them, Figure~\ref{ab.sub.1} represents learning a transfer function to enhance consensus and improve its adaptability to clients. Figure~\ref{ab.sub.2} represents learning a unified transfer function to map consensus and personal item embeddings into a common latent space to mitigate their heterogeneity.
We conduct experiments on the LastFM-2K and Douban-Book datasets and present the results in Table~\ref{tab:Ablation2}.
The results indicate that the ACE module exhibits significantly better performance than other consensus enhancement methods in both HR@10 and NDCG@10 metrics, whether on the more sparse dataset (\textit{i.e.}, LastFM-2K) and on the relatively dense dataset (\textit{i.e.}, Douban-Book). This strongly proves the superiority of the adaptive client-level consensus enhancement strategy.

\begin{figure}[t]
	\centering
	\includegraphics[width=1\linewidth]{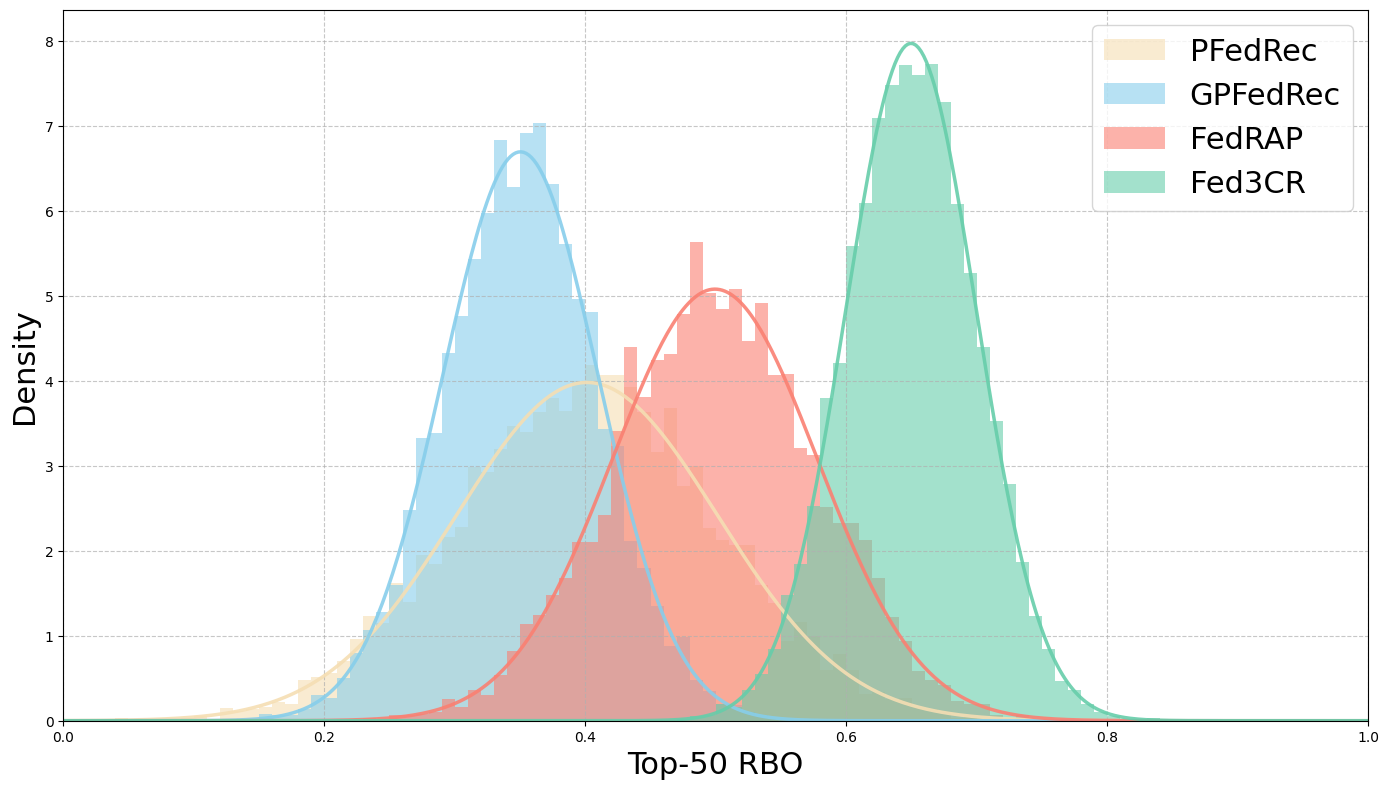}
	\caption{Top-50 Rank Biased Overlap result on MovieLens-1M dataset, where a larger RBO indicates a more consistent preference sequence, leading to better semantic consistency. We normalize the histogram area so that its height represents the probability density. See details in Sec. \ref{SEC-corr1}.}
	\label{fig:consensus}

\end{figure} 

\begin{figure}[t]
    \centering
    \subfloat[{Fed3CR ($\mathcal{L}_o$)}\label{gamma.sub.1}]{

        \includegraphics[width=3.03cm,height = 3.03cm]{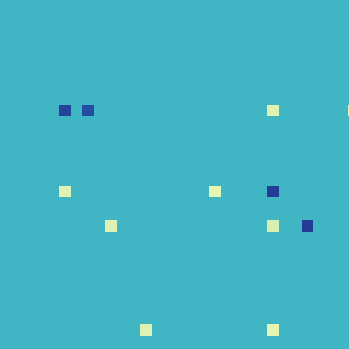}
        }
        \qquad
    \subfloat[{Fed3CR (L2)}\label{gamma.sub.2}]{
        
        \includegraphics[width=3.03cm,height = 3.03cm]{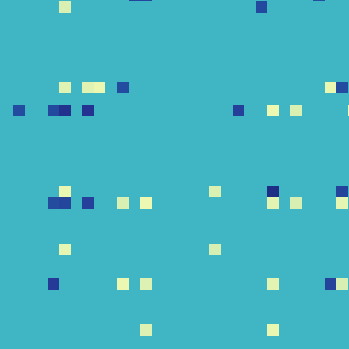}
    }
\caption{Correlation matrix $\mathcal{E}$ of client 22 on LastFM-2K dataset. $\mathcal{L}_o$ and L2 represent our proposed complementarity constraint and the L2 regularization used by FedRAP, respectively. For a more straightforward presentation, we zero out the values in [-0.003, 0.003], the highlighted parts are more relevant features. The more highlighted features there are, the weaker the complementarity. See details in Sec. \ref{SEC-corr2}.}
\label{fig:corr1}
\end{figure}

\subsection{Consistency and Complementarity (\textbf{\textit{RQ3}})}
\label{SEC-corr}

We conduct experiments to demonstrate the consistency and the complementarity of Fed3CR on MovieLens-1M and LastFM-2K datasets, respectively.

\subsubsection{Consistency}
\label{SEC-corr1}
To verify whether the proposed method can improve the semantic consistency between two views in the client, we show the Top-50 RBO results of the consensus-based FedRec methods on the MovieLens-1M dataset in Figure~\ref{fig:consensus}. We normalize the histogram area so that its height represents the probability density.
The results indicate that Fed3CR can effectively improve the semantic consistency of the enhanced consensus and personal item embeddings, which is conducive to better utility of consensus when making personalized recommendations.
Besides, combining the results in Table~\ref{tab:my_label}, it is benefit to maintain a certain consistency between the two views.

\subsubsection{Complementarity}
\label{SEC-corr2}

Figure~\ref{fig:corr1} visualizes the correlation matrices of global and personal item embedding tables of client 22 by heatmaps, with our proposed complementarity constraint $\mathcal{L}_o$ and FedRAP's L2 regularization, respectively.
For a more straightforward presentation, we zero out the values in [-0.003, 0.003], the highlighted parts are more relevant features.
Compared to L2 regularization, $\mathcal{L}_o$ can obtain more irrelevant representations. This implies that our proposed complementarity constraints can decouple global embeddings and personal item embeddings more efficiently and improve the representation of fused embeddings.

\begin{figure}
    \centering
    \vspace{-0.3cm}
    \subfloat[{H@10 on LastFM-2K}\label{ag.sub.1}]{
        
        \includegraphics[width=4.04cm,height = 3.03cm]{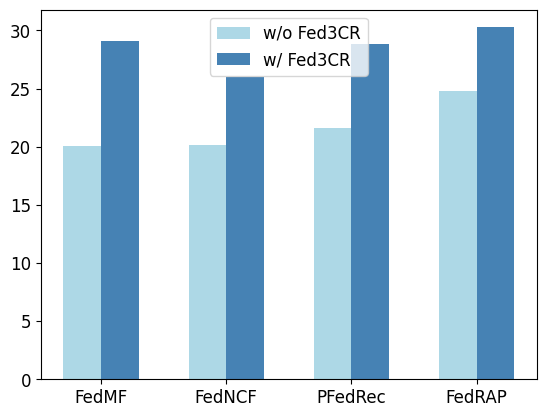}
        }
    \subfloat[{N@10 on LastFM-2K}\label{ag.sub.2}]{
        
        \includegraphics[width=4.04cm,height = 3.03cm]{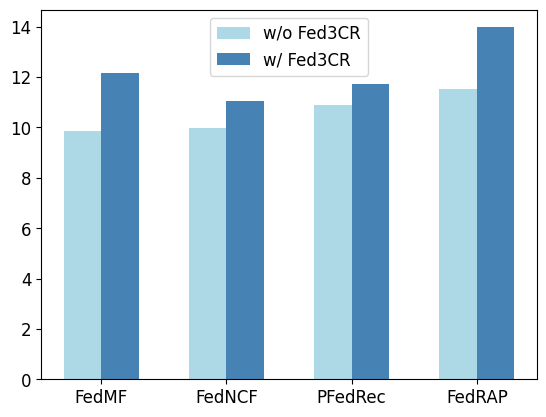}
    }
    \\
    \subfloat[{H@10 on MovieLens-1M}]{
    \label{ag.sub.3}
    \includegraphics[width=4.04cm,height = 3.03cm]{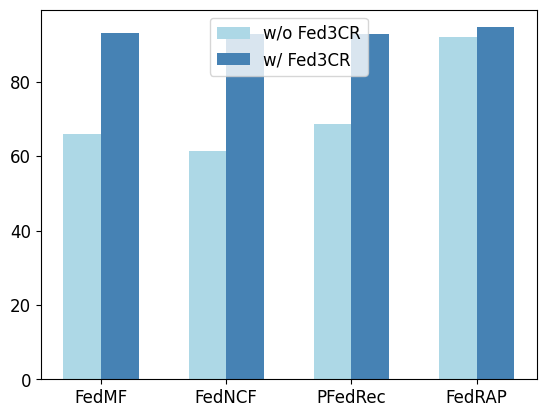}
    }
\subfloat[{N@10 on MovieLens-1M}]{
    \label{ag.sub.4}
    \includegraphics[width=4.04cm,height = 3.03cm]{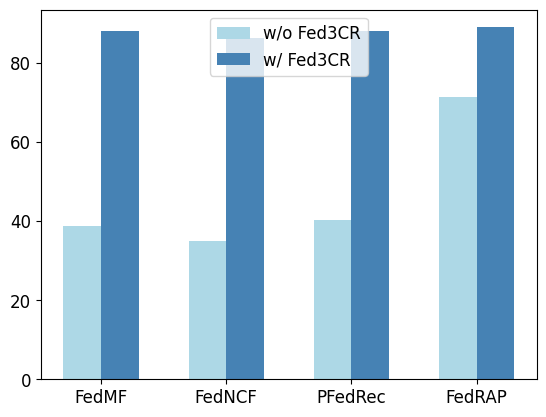}
}

\caption{Performance improvement for integrating our consensus enhancement mechanism to four FedRec baseline methods on LastFM-2K and MovieLens-1M datasets.  H@10 and N@10 indecate HR@10 and NDCG@10, respectively. See details in Sec.~\ref{sec-agnostic}.}

\label{fig:agnostic_LastFM}

\end{figure}

\subsection{Enhance Federated Recommendation Methods with Fed3CR Mechanism (\textbf{\textit{RQ4}})}
\label{sec-agnostic}
This paper proposes an efficient consensus enhancement mechanism that can be easily integrated into other FedRec methods.
We take FedMF, FedNCF PFedRec and FedRAP as examples to verify the efficacy of the consensus enhancement mechanism, and show the improvements under the sample setting on the LastFM-2K and MovieLens-1M datasets in Figure~\ref{fig:agnostic_LastFM}.
Since FedMF and FedNCF lack the consensus utilization, we treat the global item embedding tables downloaded from the server as consensus.
Evidently, the proposed consensus enhancement mechanism improves the performance of the four baselines. This indicates that introducing the consensus enhancement mechanism improves the retrieval ability of the baselines in the Top-10 recommendation task.
Meanwhile, the increased efficiency of consensus utilization enables the client to learn fine-grained relationships among items and improves the personalized recommendation performance.
Noticing that the mechanism generally improves more on MovieLens-1M than LastFM-2K, which is due to the fact that the MovieLens dataset is more dense, allowing the model to better enhance consensus and achieve better personalization.

\begin{figure*}
    \centering
    
    \subfloat[\textbf{$\beta_a$}\label{hypa.sub.1}]{
        
        \includegraphics[width=0.33\textwidth]{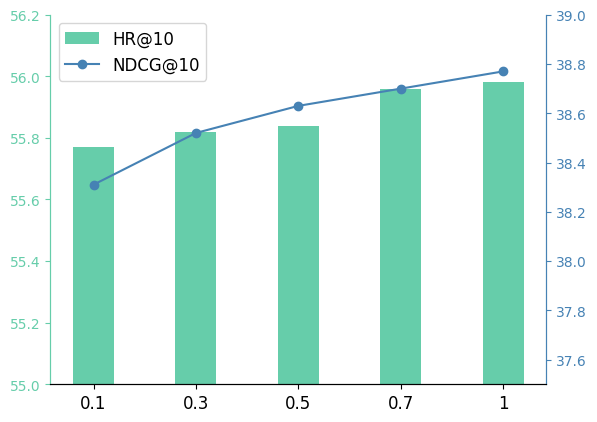}
        }
        
    \subfloat[\textbf{$\beta_o$}\label{hypa.sub.2}]{
        
        \includegraphics[width=0.33\textwidth]{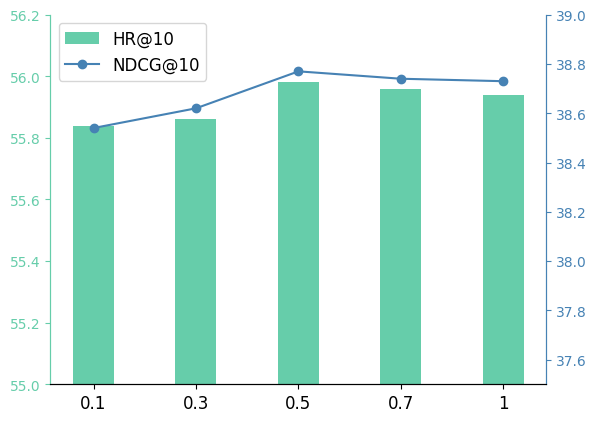}
    }
    
    \subfloat[layers of $\mathcal{F}_i(\theta_i)$\label{hypa.sub.4}]{
        
        \includegraphics[width=0.33\textwidth]{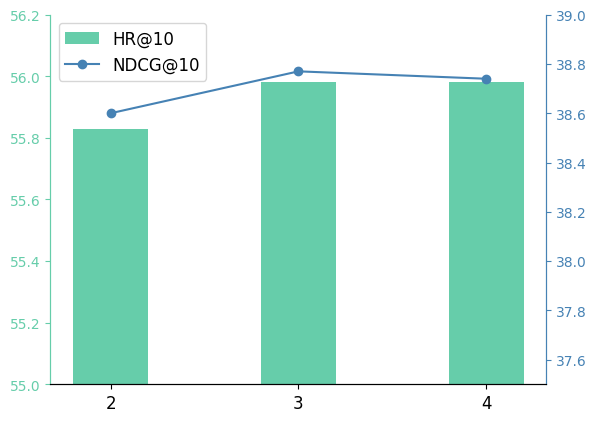}
    }
\caption{Influence of different hyperparameter on Douban-Book dataset. See details in Sec. \ref{appendix-hyper}.}
\label{fig:hyper} 

\end{figure*}

\subsection{Hyperparameter Sensitivity Analysis (\textit{RQ6})}
\label{appendix-hyper}

In this section, we study the impact of three key hyperparameters of Fed3CR on LastFM-2K dataset: the hyperparameters $\beta_a$ and $\beta_o$ in C2O  and the number of layers in $\mathcal{F}_i(\theta_i)$.

\subsubsection{Effect of $\beta_a$ for consistency}

As shown in Figure~\ref{hypa.sub.1}, increasing $\beta_a$ helps improve performance, illustrating the benefits of maintaining consistency in global and personal item embedding tables.
We observe greater gains in NDCG@10 from improving $\beta_a$, suggesting that maintaining consistency between the two views is more conducive to learning fine-grained relationships between user preferences.

\subsubsection{Effect of $\beta_o$ for complementarity}

As shown in Figure~\ref{hypa.sub.2}, a too small $\beta_o$ leads to insufficient decouple between global and personal item embedding tables. While a too large $\beta_o$ causes the model to overfocus on the differences between the two views, leading to overfitting and feature degradation.

\subsubsection{Effect of the number of layers in $\mathcal{F}_i(\theta_i)$}
We set $\mathcal{F}_i(\theta_i)$ to be fully connected networks with layers $\left[2,3,4\right]$, respectively. As shown in Figure~\ref{hypa.sub.4}, it indicates that too few layers causes the model cannot adequately learn the complex relationships between global and personal item embeddings, and too many layers can lead to overfitting. Moreover, too many layers can lead to increased communication overhead on the client side.

\section{Conclusion}
Due to the inherent sparsity and high uniformity of interaction data from massive-scale clients, FedRec faces challenges of inefficient consensus utilization and loss of personalization.
In this article, we first give a detailed explanation of the consensus degradation phenomenon in personalized FedRec and conduct some mathematical discussions.

Based on this, we propose Fed3CR, a federated consistency- and complementarity-aware consensus-enhanced recommendation approach.
Specifically, it adaptively enhances the specific information in the consensus for each client by learning the relationship between the global and personal item views, thereby improving the utilization efficiency of consensus.
Meanwhile, in order to decouple and fuse the two views more efficiently, we define consistency and complementarity between the two views and propose corresponding strategies to systematically promote them.
The experimental results fully demonstrate the superiority of the proposed Fed3CR, which effectively improves the performance of the existing consensus-based personalized FedRec.

Most importantly, our proposed Fed3CR can be integrated with other FedRec methods to improve their performance.
Overall, Fed3CR makes FedRec more competitive, providing strong technical support for scenarios that require recommendations while protecting user privacy.

\bibliographystyle{IEEEtran}
\bibliography{bare_jrnl_new_sample4}

\vfill

\end{document}